\documentclass[prd,showpacs,10pt]{revtex4}
\usepackage{graphicx}% Include figure files

\usepackage{bm}% bold math

\def\l4{l_{\rm 4}}
\def\planck5{l_{\rm 5}}

\def\AdS5{\rm AdS_5}

\begin{document}

\title{Scalar Field Cosmologies 
Hidden Within the Nonlinear Schr\"odinger Equation}
\author{James E. Lidsey}
\affiliation{Astronomy Unit, School of Physics and Astronomy,   
Queen Mary University of London, Mile End Road, LONDON, E1 4NS, U.K.,   
J.E.Lidsey@qmul.ac.uk}

%\date{\today}

\begin{abstract}
The nonlinear, cubic Schr\"odinger (NLS) equation has numerous 
physical applications, but in general is very difficult to solve. 
Nonetheless, under certain circumstances parameters quantifying the
width, momentum and energy of the wavefunction evolve under a closed 
set of ordinary differential equations. It is shown that for the 
case of the radial, two dimensional NLS equation, such evolution equations 
may be mapped directly onto the cosmological Friedmann equations for a 
spatially flat and isotropic universe sourced by a self-interacting 
scalar field and a barotropic perfect fluid. Consequently, analytical 
techniques that have been developed to study the dynamics of such cosmological
models may be applied to gain insight into aspects of 
nonlinear quantum mechanics. In this 
paper, the Hamilton-Jacobi formalism of the Friedmann equations, where the
scalar field is viewed as the dynamical variable, is developed within this
context. Algorithms for finding exact solutions are presented 
and the scaling solutions determined. 
A form-invariance of the wavefunction evolution equations is
identified. The analysis has direct applications to anisotropic 
Bose-Einstein condensation. The Ermakov-Pinney equation 
plays a central role in establishing the correspondence between the
quantum-mechanical and gravitational systems. 
\end{abstract}
 
\vskip 1pc 
\maketitle

\section{Introduction}

\label{intro}

\def\theequation{\arabic{equation}}

In the absence of a consistent, non-perturbative theory of quantum gravity, 
it is important to establish 
connections between quantum mechanical and gravitational systems. 
One route towards such a goal is to 
uncover direct links between the underlying 
equations of motion for different models. 

From the quantum mechanical perspective,   
an important family of equations are nonlinear Schr\"odinger (NLS) equations, 
which arise for example in the fields of Bose-Einstein condensation
\cite{G,P,volovik1,volovik2,frant} and 
quantum optics \cite{hasegawa}. 
In $d$ spatial dimensions, the NLS equation with a ${\rm U}(1)$-invariant 
interaction term has the generic form: 
\begin{equation}
\label{genNLS}
i \frac{\partial u}{\partial \tau} = -\frac{1}{2} \nabla^2_d u 
+ V( \mathbf{r}, \tau ) u + \nu (\tau ) | u|^{2(m-1)} u \, ,
\end{equation}
where $\nabla^2_d$ is the $d$-dimensional spatial Laplacian,
$V (\mathbf{r}, \tau)$ 
represents a time-dependent potential, $\nu (\tau)$ is a 
time-dependent coupling parameter and $m$ is a constant. 
Due to the nonlinearities involved, 
it is difficult (if not impossible) to find exact solutions to such 
an equation. Nonetheless, insight into 
the properties of the wavefunction can be gained by employing the `moment
method' \cite{moment,moment1,moment2,moment3,moment4}. 
In this approach, physical quantities 
such as the width, momentum and energy 
of the quantum system can be defined in terms of integral relations 
involving the 
wavefunction. These `moments' satisfy a set of coupled, linear, first-order 
ordinary differential equations (ODEs) that derive from the NLS 
equation. 

On the other hand, cosmology provides a natural environment 
for the study of gravitational physics. 
The Friedmann equations (the time-time and 
space-space components of the Einstein
field equations) are a cornerstone of modern cosmology and 
relate the expansion of the universe to the total energy density
contained within it. Of particular interest are spatially isotropic 
cosmological 
models that are sourced by a mixture of self-interacting scalar fields and 
perfect (barotropic) fluids. 
Such models play an important role in the inflationary
scenario of the very early universe and are  
a leading candidate for the origin of the 
dark energy that dominates the universe at the current epoch (for reviews, 
see, e.g., \cite{lidsey,lr,cst,clifton}). Scalar fields also 
arise ubiquitously in 
cosmologies inspired by superstring/M-theory (see Refs. 
\cite{pbb1,pbb2} and references therein).

The purpose of the present paper is to uncover a connection 
between the dynamics associated with such cosmological models and the 
radially symmetric, cubic NLS equation. 
The link is made possible due to the central role played by 
a nonlinear, second-order ODE of the form 
\begin{equation}
\label{EPintro}
\frac{d^2X(\tau)}{d\tau^2} + \lambda (\tau) X(\tau) = \frac{Q}{X^3(\tau)}
\, ,
\end{equation}
where $X(\tau )$ and $\lambda (\tau )$ are functions 
of the independent variable
$\tau$ and $Q$ is a constant. Eq. (\ref{EPintro}) is known as the 
Ermakov-Pinney (EP) equation \cite{ermakov,pinney,milne}. 
It plays a fundamental role in diverse branches of 
mathematical and theoretical physics, ranging from nonlinear optics, nonlinear
elasticity, molecular structures, Bose-Einstein condensation, cosmology and 
quantum cosmology. (A brief survey of the algebraic 
properties and physical applications of the 
EP equation can be found in Refs. \cite{EPsurvey,EPsurvey1}.) 
Its role in cosmology has been discussed by a number of 
authors 
\cite{hawklid,williams,lidseycondensate,williams1,herring,ambroise}. 

For the classes of quantum and cosmological 
models we consider, the physical 
interpretations of the dependent and independent variables in 
Eq. (\ref{EPintro}) differ. This allows a dictionary 
relating different variables to be established.  
It is then found that the 
equations of motion for the quantum and gravitational models are equivalent, 
in the sense that there exists a one-to-one correspondence that maps 
the cosmological Friedmann equations onto 
the evolution equations of the wavefunction moments (and vice-versa). 
In other words, the field equations for the two systems are 
formally interchangeable. 

Consequently, analytical techniques that have been developed previously for
studying the early- and late-time dynamics of the universe
can be employed to gain insight into the nature 
of the wavefunction of the NLS equation. 
Conversely, nonlinear quantum dynamics can be employed to 
develop an alternative formulation of the Einstein equations. 

We focus on applying the 
`Hamilton-Jacobi' (HJ) formalism of the cosmological field equations 
\cite{HJ,HJ1,HJ2,HJ3,HJ4,lidsey*,lidsey1} to the 
moment equations of motion. 
In this framework, the scalar field is viewed as the 
independent dynamical variable rather than cosmic time. This allows the
second-order Friedmann equations to be expressed explicitly as a first-order
system. Within the context of 
the present-work, we present 
an alternative way of expressing the moment evolution equations 
in terms of a coupled, first-order system of ODEs. A form-invariance  
of the ODE's is uncovered, in the sense that the set of ODEs 
is invariant under a non-local transformation group of the variables
\cite{chimento}.

The paper is organized as follows. Sections \ref{cosmic} and 
\ref{bose-einstein}  briefly 
review scalar field cosmology and the moment method, respectively. 
The dictionary between different cosmological and quantum-mechanical models 
is established in Section \ref{dictionary}. The HJ approach to 
cosmology is applied in Section \ref{HJNLS} and the underlying form-invariance 
of the differential equations made manifest. 
In Section \ref{algorithms}, we present algorithms for finding exact solutions 
and determine the analogues of cosmological scaling solutions. 
We conclude in Section \ref{conclusion} with a discussion. 

Unless otherwise stated, 
units are chosen such that $\hbar = c =1$ and  $m_P = \sqrt{8\pi}$, 
where $m_P$ is the Planck mass. 

\section{Cosmological Friedmann Equations}

\label{cosmic}

\def\theequation{\arabic{equation}}

The field equations for the spatially flat, isotropic, 
Friedmann-Robertson-Walker (FRW) 
universe sourced by a barotropic perfect fluid and a 
minimally coupled scalar field $\phi$ self-interacting 
through a potential $W(\phi )$ are given by the Friedmann equation
\begin{equation}
\label{friedmann} 
3 H^2 =  
\frac{1}{2} \left( \frac{d\phi}{dt} \right)^2  +W(\phi)
+ \frac{D}{a^n} 
\end{equation}
and the scalar field equation of motion: 
\begin{equation}
\label{scalareom}
\frac{d^2\phi}{dt^2} + 3H \frac{d\phi}{dt} + \frac{dW}{d\phi} =0 \, ,
\end{equation}
where $H \equiv d \ln a /dt$ is the Hubble parameter of the scale factor
$a(t)$, $t$ denotes cosmic time,
$\rho_{\rm mat} = Da^{-n}$ is the energy density of the fluid 
with an equation of state $p_{\rm mat} =[(n-3)/3]\rho_{\rm mat}$ and  
$\{ D, n\}$ are constants. Within the context of Einstein gravity, causality 
implies the bounds $D>0$ and $0 \le n \le 6$. For $n <0$ (corresponding to
`phantom' matter), the null energy condition is violated, $\rho_{\rm mat} 
+p_{\rm mat} <0$. For $n>6$ (corresponding to ultra-stiff matter), the 
speed of sound in the fluid exceeds the speed of light, $c_{\rm s} =
\sqrt{dp/d\rho} >1$. For $D<0$, the weak energy condition is violated, 
$\rho_{\rm mat} <0$. However, since we are interested in establishing 
correspondences between different
systems, we leave the fluid parameters 
unspecified in what follows. 
Indeed, it is worth noting that quantum cosmological 
considerations typically lead to corrections to the standard Friedmann 
equation, where $D$ is effectively negative. This is the case for example in 
braneworld \cite{sahni}
and loop quantum cosmology \cite{param} scenarios, 
where a term proportional to $-\rho_{\rm mat}^2$ 
generically arises. The above equations also apply to spatially 
curved FRW universes $(n=2)$ and the anisotropic, Bianchi I cosmology 
$(n=6)$. 

Differentiating the Friedmann equation (\ref{friedmann}) and substituting for 
Eq. (\ref{scalareom}) yields the Raychaudhuri equation:  
\begin{equation}
\label{ray}
\frac{1}{a} \frac{d^2a}{dt^2} - \frac{1}{a^2} \left( \frac{da}{dt} 
\right)^2 = -\frac{1}{2} \left[ \left( \frac{d\phi}{dt}  \right)^2 + 
\frac{nD}{3a^n} \right] \, .
\end{equation}
We may rewrite this equation by defining an effective scale factor:
\begin{equation}
\label{defb}
b \equiv a^{n/2}
\end{equation}
and a new time parameter through the relation:
\begin{equation}
\label{deftau}
\frac{d}{dt} \equiv a^{n/2} \frac{d}{d\tau} \, .
\end{equation}
It follows that Eq. (\ref{ray}) transforms into the EP equation
\cite{hawklid}:
\begin{equation}
\label{cosmicEP}
\frac{d^2b}{d\tau^2} + \frac{n}{4} \left( \frac{d\phi}{d\tau} 
\right)^2b = -\frac{Dn^2}{12} \frac{1}{b^3} \, .
\end{equation}
The inclusion of a perfect fluid in the energy-momentum tensor $(D\ne 0)$ 
results in the nonlinear cubic term on the right-hand side of 
Eq. (\ref{cosmicEP}).  

Finally, we note for future reference that the scalar field equation 
(\ref{scalareom}) can be written 
as the first-order differential equation:  
\begin{equation}
\label{dotrho}
\frac{d\rho_{\phi}}{dt} = - 3 H \left( \frac{d\phi}{dt} \right)^2 \, ,
\end{equation}
where $\rho_{\phi} \equiv \frac{1}{2} (d\phi/dt)^2 +W(\phi)$ 
represents the energy density of the field.  

\section{Nonlinear Schr\"odinger Equation and the Moment Method}

\label{bose-einstein}

\def\theequation{\arabic{equation}}

In this paper, we focus on
the two-dimensional, radially symmetric form of Eq. (\ref{genNLS}) 
with a cubic interaction term $(m=2)$ and 
an harmonic potential given by $V(r, \tau) = \lambda (\tau ) r^2/2$, 
where $\lambda (\tau )$ is a time-dependent frequency. We further assume that  
the parameter $\nu$ is constant. In this case, Eq. (\ref{genNLS})
reduces to
\begin{equation}
\label{2dNLS}
i \frac{\partial u}{\partial \tau} = - \frac{1}{2r} 
\frac{\partial}{\partial r} \left( r \frac{\partial u}{\partial r} \right)
+ \frac{1}{2} \lambda (\tau ) r^2 u + \nu  |u|^2  u  \, .
\end{equation}
We refer to the independent variable, $\tau$, 
as `laboratory time' throughout the discussion.
%Eq. (\ref{genNLS}) is referred to as the 
%focusing (defocusing) NLS equation for $\nu < 0$ $(\nu > 0)$.
Within the context of Bose-Einstein condensation 
\cite{volovik1,volovik2,frant}, Eq. (\ref{2dNLS}) 
is known as the Gross-Pitaevskii equation \cite{G,P}. It determines the 
dynamics when the condensate is tightly bound along one direction by the 
trapping potential $V(r,\tau )$, in which case
a quasi-two-dimensional system is established. 
The parameter $\nu$ is proportional to the $s$-wave scattering length 
that quantifies the strength of the atomic interactions. 

The moments associated with the wavefunction are defined by 
the integral quantities \cite{moment1}: 
\begin{eqnarray}
\label{I1}
I_1 (\tau ) & = & \int |u|^2 d^2x  \, ,
\\
\label{I2} 
I_2 ( \tau ) & = & \int |u|^2 r^2 d^2x  \, ,
\\
\label{I3}
I_3 (\tau ) & = & i \int \left( u \frac{\partial u^*}{\partial r}
- u^* \frac{\partial u}{\partial r} \right) r d^2x \, ,
\\
\label{I4}
I_4 (\tau ) & = & \frac{1}{2} \int \left( \left| \nabla u \right|^2 + 
\nu |u|^4 \right) d^2x  \, ,
\end{eqnarray}
respectively, where $d^2x = rdrd\theta$ and integration 
over the angular coordinate yields a factor of $2\pi$. 
A star denotes complex conjugate. 
The moments $I_j$ $(j=2,3,4)$ 
quantify the width, radial momentum
and energy of the quantum configuration, respectively.  

Differentiating Eqs. (\ref{I1})-(\ref{I4}) with respect to 
laboratory time, substituting
Eq. (\ref{2dNLS}) 
and its complex conjugate, and integrating by parts a number of times
yields (after some algebra) a set of 
coupled, first-order, linear ODEs \cite{moment1}: 
\begin{eqnarray}
\label{I1ode}
\frac{dI_1}{d \tau} & = & 0  \, 
\\
\label{I2ode}
\frac{dI_2}{d\tau} & = & I_3  \, ,
\\
\label{I3ode}
\frac{dI_3}{d\tau} & = & -2 \lambda (\tau ) I_2 +4I_4  \, ,
\\
\label{I4ode}
\frac{dI_4}{d\tau } & = & -\frac{1}{2} \lambda (\tau ) I_3 \, .
\end{eqnarray}

Eqs. (\ref{I1ode})-(\ref{I4ode}) represent a set of closed evolution laws and  
admit an invariant under time evolution: 
\begin{equation}
\label{defQ}
Q = 2 I_2 I_4 - \frac{1}{4} I_3^2  \, ,
\end{equation}
as may be verified by direct differentiation and substitution. 
This constant may be negative for sufficiently negative $\nu$. 
Given the constraint equation (\ref{defQ}), 
the system (\ref{I1ode})-(\ref{I4ode}) can be reduced to a single, second-order
differential equation
\begin{equation}
\label{I2equation}
\frac{d^2I_2}{d\tau^2} - \frac{1}{2I_2} \left( \frac{dI_2}{d\tau} \right)^2 +2
\lambda (\tau) I_2 = \frac{2Q}{I_2}
\end{equation}
and Eq. (\ref{I2equation}) transforms into the EP equation \cite{moment}
\begin{equation}
\label{NLSEP}
\frac{d^2 X}{d\tau^2} + \lambda (\tau ) X = 
\frac{Q}{X^3}  \, ,
\end{equation}
where $X (\tau )\equiv I_2^{1/2}$ determines the width of the 
wavepacket. 
Hence, solutions to the moment equations
(\ref{I2ode})-(\ref{I4ode}) follow directly from solutions 
to the EP equation (\ref{NLSEP}) (and vice-versa).   
Since the $L^2$-norm of the wavefunction is conserved, the amplitude of the 
wavefunction is approximately $A \sim 1/I_2$, i.e., the inverse square of the 
width of the wavepacket. 

\section{A Dictionary Between Quantum-Mechanical and Cosmological Systems}

\label{dictionary}

\def\theequation{\arabic{equation}}

A direct comparison between the `cosmological' EP equation (\ref{cosmicEP})
and the `Schr\"odinger' EP equation (\ref{NLSEP}) 
immediately suggests that a formal correspondence may be established between 
appropriate cosmic and wavefunction parameters. Specifically, we identify  
\begin{eqnarray}
\label{formal1}
I_2 (\tau (t)) & = & a^n (t)   \, ,
\\
\label{formal2}
\lambda (\tau (t) ) & = &
\frac{n}{4} \left( \frac{d\phi}{d\tau} 
\right)^2  \, ,
\\
\label{formal3}
Q & = & -\frac{Dn^2}{12} \, ,
\end{eqnarray}
where the relation between cosmic and laboratory times is given by integrating 
Eq. (\ref{deftau}): 
\begin{equation}
\label{formaldeftau}
t (\tau ) = \int^{\tau} \frac{d\tau}{\sqrt{I_2(\tau)}} , \qquad 
\tau (t) =\int^t dt \, a^{n/2} (t)  \, .
\end{equation}

It then follows from Eqs.  (\ref{deftau}) and (\ref{I2ode}) that 
\begin{equation}
\label{formal4}
I_3 (\tau (t)) = na^{n/2} (t) H (t)  
\end{equation}
and further insight may be gained if we identify the moment $I_4$ with the 
energy density of  the scalar field such that
\begin{equation}
\label{formal5} 
I_4 (\tau (t)) = \frac{n^2}{24}  \rho_{\phi} (t) \, .
\end{equation}
It may now be verified after substitution of Eqs. (\ref{formal1}), 
(\ref{formal4}) and (\ref{formal5}) into the constraint equation
(\ref{defQ}) that the latter corresponds {\em precisely}
to the Friedmann equation (\ref{friedmann}). Furthermore, 
some straightforward algebra implies that when the above 
correspondence is applied, the equation of motion for the moment
$I_4$, Eq. (\ref{I4}), transforms 
directly into the scalar field equation (\ref{dotrho}) and that 
the corresponding equation for $I_3$, Eq. (\ref{I3}),
transforms directly into the Raychaudhuri equation 
(\ref{ray}). The evolution equation for $I_2$, Eq. (\ref{I2}), 
corresponds to the definition of the Hubble parameter, $H \equiv 
d\ln a /dt$. 

From a dynamical point of view, therefore, the gravitational Friedmann
equations are equivalent to the coupled evolution laws of the wavefunction 
moments. In this sense, classical cosmic dynamics is contained 
within the quantum-mechanical dynamics of the nonlinear, cubic 
Schr\"odinger equation (and vice-versa). In other words, 
the cosmological Friedmann equations 
can be directly transformed into the equations of motion for the wavefunction 
moments after appropriate redefinitions of the dependent 
and independent variables.  
The correspondence is summarized in Tables I and II. 

\begin{table}
\begin{center}
\begin{tabular}{|c|c|}
\hline 
Cosmological     & Wavefunction        \\
Parameter        & Moments           \\ 
\hline
$a$              & $I_2^{1/n}$         \\
$H$              & $I_3/(n I_2^{1/2})$ \\
$\rho_{\phi}$           & $24I_4/n^2$         \\
$t$              & $\int d\tau I_2^{-1/2}$ \\
$\Omega_{\phi}$  & $8I_2I_4/I_3^2$     \\
\hline
\end{tabular}
\end{center}
\footnotesize{\hspace*{.3in} 
\noindent
Table I: 
A dictionary between the cosmological and wavefunction variables. 
The cosmological scale factor, Hubble parameter and scalar field energy 
density are related to the width, momentum and energy of the quantum 
configuration, respectively. Cosmic time, $t$, and laboratory time, $\tau$,
are related by Eq. (\ref{formaldeftau}). 
The fraction of the scalar field energy density 
relative to the total cosmic density can be expressed as a
combination of all three wavefunction moments.}
\end{table}

\begin{table}
\begin{center}
\begin{tabular}{|c|c|}
\hline 
Cosmological          & Moment   \\
Field Equation        & Equation of Motion      \\ 
\hline
Friedmann Equation (\ref{friedmann}) & Constraint Equation (\ref{defQ})    \\
Scalar Field Equation (\ref{dotrho}) & $I_4$ Equation (\ref{I4ode})  \\
Raychaudhuri Equation (\ref{ray})    & $I_3$ Equation (\ref{I3ode}) \\
Hubble Parameter            &  $I_2$ Equation (\ref{I2ode}) \\
\hline
\end{tabular}
\end{center}
\footnotesize{\hspace*{.3in} Table II: A direct link between 
the Einstein field
equations for a spatially isotropic and flat universe sourced by a
self-interacting scalar field and the evolution laws
for the wavefunction moments of the nonlinear Schr\"odinger equation.
The equation of motion for $I_2$ is equivalent to the definition of the Hubble 
parameter.}
\label{table2}
\end{table}

From the cosmological perspective, an important physical parameter is the
fraction of the scalar field energy density relative to the total 
energy density. This is quantified in terms of the ratio
$\Omega_{\phi} \equiv \rho_{\phi}/(3H^2)$. 
In the present context, $\Omega_{\phi}$
can be related to a combination of the three moments: 
\begin{equation}
\label{phiOmega}
\Omega_{\phi} = 8 \frac{I_2I_4}{I_3^2}, 
\qquad I_3^2 = - \frac{4Q}{1- \Omega_{\phi}}  \, .
\end{equation}

The above discussion is interesting because it implies that the
techniques that have been developed for analyzing the dynamics 
of scalar field and perfect fluid cosmologies in a variety of different 
settings can be carried over to study the dynamics of 
the wavefunction of the cubic Schr\"odinger equation (and vice-versa). 
This is the topic of the next Section. 

\section{Hamilton-Jacobi Formulation and Form-Invariance 
of the Moment Evolution Equations}

\label{HJNLS}

\def\theequation{\arabic{equation}}

The cosmological field equations 
(\ref{friedmann})-(\ref{ray}) can be written in an alternative
first-order form by 
interpreting the scalar field as the effective dynamical variable 
of the system \cite{lidsey1}. 
Assuming the field to be evolving monotonically with cosmic time 
(as is the case, 
for example, for inflationary models where the field rolls slowly 
down its potential), Eq. (\ref{scalareom}) can be expressed as
\begin{equation}
\label{phiscalareom}
\frac{d\rho_{\phi}}{d\phi} =-3H \frac{d\phi}{dt}  \, .
\end{equation}
From the definition of the Hubble parameter, it then follows that 
$3nH^2=-\rho_{\phi}' \chi' /\chi$, where 
$\chi \equiv a^n$ and a prime denotes $d/d\phi$ in this and following
Sections. Substituting this expression into Eq. (\ref{friedmann}) implies that  
the Friedmann equation can be rewritten in the 
`Hamilton-Jacobi' form:
\begin{equation}
\label{phifriedmmann}
\frac{d\rho_{\phi} (\phi)}{d\phi} \frac{d \chi (\phi)}{d\phi} +n 
\rho_{\phi} (\phi) \chi (\phi) =-nD  \, .
\end{equation}

On the other hand, 
invoking the dictionary summarized in Table I implies that
the scalar field equation of motion (\ref{phiscalareom}) can 
also be expressed as  
\begin{equation}
\label{I4phiode}
\frac{d I_4}{d\phi} =-\frac{n}{8} I_3 \frac{d \phi}{d \tau} 
\end{equation}
and multiplying both sides of this expression 
by $dI_2/d\phi$ yields the relation  
\begin{equation}
\label{invariantequation}
I_3^2 = - \frac{8}{n} \frac{dI_2}{d\phi} \frac{dI_4}{d\phi} \, ,
\end{equation}
where we have employed Eq. (\ref{I2ode}). 
As a result, Eq. (\ref{defQ}) transforms to 
\begin{equation}
\label{HJQconstraint}
\frac{dI_2}{d\phi} \frac{dI_4}{d\phi} + n I_2I_4 = \frac{nQ}{2} \, .
\end{equation}

We have therefore rewritten the moment evolution equations 
(\ref{I3ode})-(\ref{defQ}) in the form 
(\ref{I4phiode})-(\ref{HJQconstraint}), where 
the new independent variable is related to laboratory time 
by integrating Eq. (\ref{formal2}):   
\begin{equation}
\label{phitau}
\phi (\tau) 
= \frac{2}{\sqrt{n}}\int^{\tau} d \tau \sqrt{\lambda(\tau)}  \, .
\end{equation} 
We choose the positive root without loss of generality, since the negative root
corresponds to a time-reversal. Assuming the scalar field 
to be a real-valued 
variable implies $\lambda < 0 \, {\rm iff} \, n<0$. 

Eq. (\ref{HJQconstraint}) is a key result of the paper 
and is interesting for a number of reasons.  
It is immediately apparent 
that is is invariant under the simultaneous transformation,
$I_2 (\phi ) \longleftrightarrow I_4 (\phi )$,
that interchanges the width and energy moments 
(when both are expressed explicitly as 
functions of the variable, $\phi$). In other words, given a solution
$\{ I_2 (\phi ), I_4 (\phi )\}$, we may generate a new `dual' solution 
$\{ \tilde{I}_2 (\phi ), \tilde{I}_4 (\phi )\}$, where 
\begin{equation}
\label{duality}
\tilde{I}_2 (\phi ) = I_4 (\phi ) , \qquad \tilde{I}_4(\phi ) = I_2 (\phi )
\, .
\end{equation}
Moreover, it follows from Eq. (\ref{defQ}) that the radial 
momentum is a singlet under this transformation:  
\begin{equation}
\label{I3singlet}
\tilde{I}_3 (\phi) = I_3 (\phi) \, .
\end{equation}

Eqs. (\ref{duality}) and (\ref{I3singlet}) represent 
a `form-invariance' of the moment 
evolution equations, in the sense that 
Eqs. (\ref{invariantequation}) and (\ref{HJQconstraint}) 
preserve their analytical form 
under the transformation (\ref{duality})-(\ref{I3singlet}). 
The invariance becomes manifest 
when these equations are expressed in terms of the 
appropriate independent variable, in this case, $\phi$. 
The two solutions are not equivalent, however, since the 
potential, $\lambda (\phi)$, is not a singlet. 
The functional forms of the potential 
frequencies that generate the solution pair are 
related through Eq. (\ref{I4phiode}) and it follows that 
\begin{equation}
\label{duallambda}
\tilde{\lambda} (\phi) = \frac{4}{\lambda (\phi)} \, .
\end{equation}
 
Eq. (\ref{I4phiode}) further 
implies that the relationship between the scalar field 
and laboratory time changes in the dual solution. Labelling the new 
functional dependence by $\tilde{\tau}(\phi )$, we deduce that 
\begin{equation}
\label{dualtau}
\tilde{\tau} (\phi) = \frac{1}{2} \int^{\phi} d \phi 
\frac{I_4'}{(2I_2I_4-Q)^{1/2}} = - \frac{1}{2} \int d \tau \lambda (\tau )
\, .
\end{equation}
%This minus sign arises because the derivatives of the moments $I_2$ and $I_4$
%have different signs. 
Since the derivatives $I_2'$ and $I_4'$ have opposite signs, the 
interchange $I_2(\phi) \leftrightarrow I_4(\phi)$ 
maps an expanding configuration onto a collapsing one, and vice-versa. 
The analogous cosmological transformation would relate an expanding universe 
to a contracting one \cite{lidsey2}. In this sense,
Eqs. (\ref{duality})-(\ref{I3singlet}) 
represent an implosion/explosion duality of the quantum system.  

Given the relation (\ref{dualtau}) 
between the dual `laboratory' times $\tau$ and 
$\tilde{\tau}$, it is now straightforward to verify that 
the set of coupled, moment evolution equations (\ref{I2})-(\ref{defQ}) 
are form-invariant under 
a redefinition of the dependent and independent variables 
such that  
\begin{eqnarray}
\label{EPinvariance}
\tilde{I}_2 (\tilde{\tau} ) =  I_4(\tau ) , \qquad
\tilde{I}_3 (\tilde{\tau} ) =  I_3(\tau ) , \qquad
\tilde{I}_4 (\tilde{\tau} )  =& I_2(\tau ) , \qquad
\nonumber
\\
\tilde{\lambda} (\tilde{\tau})  = \frac{4}{\lambda (\tau )} , 
\qquad \frac{d}{d\tilde{\tau}} =  -\frac{2}{\lambda(\tau )} \frac{d}{d\tau}
\, .
\end{eqnarray}
It should be emphasized that whilst the redefinitions (\ref{EPinvariance})
leave the equation of motion for $I_3 (\tau)$ invariant, they interchange the 
evolution laws for $I_2 (\tau)$ and $I_4 (\tau)$. Hence, given a solution set 
$\{ I_j (\tau ) , \lambda (\tau)\}$, we may generate a new 
solution set $\{ \tilde{I}_j (\tilde{\tau} ) , 
\tilde{\lambda} (\tilde{\tau })\}$. This can be done 
analytically if the integral (\ref{dualtau}) can be evaluated and inverted. 

An immediate consequence of Eqs. (\ref{EPinvariance}) 
is that the EP equation (\ref{EPintro}) 
exhibits the same form-invariance when 
the dependent variable is identified as $\sqrt{I_2 (\tau)}$ and 
the moment equations of motion are satisfied. Moreover, 
the above discussion also applies when $Q=0$ and therefore to 
the corresponding LS equation.  

We employ the above `Hamilton-Jacobi' formalism for the moment evolution 
equations in the following Section to find analytical solutions of the system 
in terms of quadratures. 

\section{Algorithms for Finding Exact Analytical Solutions}

\label{algorithms}

\def\theequation{\arabic{equation}}

\subsection{General Solution in terms of Quadratures}

Eq. (\ref{HJQconstraint}) can be solved in full generality 
in terms of quadratures for either $I_2(\phi)$ or $I_4 (\phi)$ 
and their corresponding first derivatives. For 
example, the width of the configuration associated with the wavefunction is 
given in terms of its energy by
\begin{equation}
\label{I2quad}
I_2( \phi) = \exp \left[ -n \int^{\phi} d\phi \frac{I_4}{I_4'} \right]
\times \left[ \Pi_2 + \frac{nQ}{2} 
\int^{\phi} d\phi \frac{1}{I_4'} \exp \left( n 
\int^{\phi} d\phi \frac{I_4}{I'_4}
 \right) \right] \, ,
\end{equation}
where $\Pi_2$ is an arbitrary integration constant. 
A similar expression arises for the energy of the wavepacket in terms 
of its width: 
\begin{equation}
\label{I4quad}
I_4( \phi) = \exp \left[ -n \int^{\phi} d\phi \frac{I_2}{I_2'} \right]
\times \left[ \Pi_4 + \frac{nQ}{2} 
\int^{\phi} d\phi \frac{1}{I_2'} \exp \left( n 
\int^{\phi} d\phi \frac{I_2}{I'_2}
 \right) \right]
\end{equation}
for arbitrary constant $\Pi_4$.

Thus, the width of the wavepacket can in principle be determined 
parametrically if the functional form 
of $I_4(\phi)$ is specified. In this case, 
the corresponding expression for the momentum is given by  
\begin{equation}
\label{I3quad}
I_3 (\phi) = \pm 2 \sqrt{2I_2 (\phi) I_4 (\phi) -Q}
\end{equation}
and the potential frequency follows from 
Eqs. (\ref{I4phiode})-(\ref{invariantequation}):
\begin{equation}
\label{lambdaphi}
\lambda (\phi) =  -2 \frac{I_4'}{I_2'}
= \frac{4}{n} \frac{I_4'^2}{(2I_2 I_4 -Q)}  \, .
\end{equation}
Finally, laboratory time is determined parametrically by
\begin{equation}
\label{tauphi}
\tau (\phi) -\tau_0 = 
\sqrt{\frac{n}{8}} \int^{\phi} d \phi \left( -\frac{I_2'}{I_4'} \right)^{1/2}
= \frac{n}{4} \int^{\phi} d\phi 
\frac{\sqrt{2I_2I_4 -Q}}{I_4'} \, ,
\end{equation}
where we have denoted the integration constant by $\tau_0$. 

We have therefore expressed the {\em general} solution to the moment evolution 
equations (\ref{I2ode})-(\ref{I4ode}) in parametric 
form in terms of quadratures with respect to a single (to be specified) 
function, $I_4(\phi )$, and its first derivative with respect to 
the new independent variable, $\phi$. 
(The solution is general since there are three independent constants 
$\{ Q , \Pi , \tau_0 \}$ .) 

This suggests an algorithm to find exact solutions to the moment 
equations. Specify the functional form of the configuration 
energy as a function of the field
$\phi$. Integrate Eq. (\ref{I2quad}) to determine the width $I_2(\phi)$. 
The momentum $I_3(\phi )$ 
and potential $\lambda (\phi )$ follow directly from Eqs. (\ref{I3quad})
and (\ref{lambdaphi}), respectively. The dependence of this set of 
physical parameters on laboratory time then follows by integrating and 
inverting Eq. (\ref{tauphi}). 
The primary challenge in this approach is 
to successfully complete the last step in the iteration. Nonetheless, 
we present a solution to the equations of motion
by employing this method in the Appendix. 

An alternative approach is to first specify the energy density explicitly as a
function of the width, i.e., choose an appropriate functional 
dependence $I_4 = F [I_2 (\phi)]$. Eq. (\ref{HJQconstraint}) then 
reduces to an ODE of the form 
\begin{equation}
\label{I2only}
\left( \frac{dI_2}{d\phi} \right)^2 + n \frac{F}{F^*} I_2 = 
\frac{nQ}{2} \frac{1}{F^*}  \, ,
\end{equation}
where $F^* \equiv dF/dI_2$. For cases where Eq. (\ref{I2only}) is solvable, 
the algorithm then proceeds as above. 

It should be emphasized that even in cases where a solution in terms of 
laboratory or cosmic time can not be deduced analytically, 
important information regarding the dynamics of a given system can still 
be inferred in terms of the independent variable, $\phi$. To illustrate this
point, let us consider the {\em ansatz}: 
\begin{equation}
\label{I2I4ansatz}
I_4 (\phi) = C I_2^{-\beta} (\phi)  \, ,
\end{equation}
where $\{ C , \beta \}$ are arbitrary constants and let us 
further assume that $n=1$ (for algebraic simplicity) and that 
$Q>0$ (a similar analysis can be performed when $Q<0$). 
Eq. (\ref{I2ode}) reduces 
to 
\begin{equation}
\label{I2I4ansatzode}
\left( \frac{dI_2}{d\phi} \right)^2 = \frac{1}{\beta} \left( I_2^2 -
\frac{Q}{2C} I_2^{1+\beta} \right)
\end{equation}
and defining a new dependent variable, $z$, such that
\begin{equation}
\label{defz}
I_2 \equiv \left( \frac{2C}{Q} \right)^{1/(\beta -1)} 
\left[ {\rm sech} \, z \right]^{2/(\beta -1)}
\end{equation}
then transforms Eq. (\ref{I2I4ansatzode}) into the trivial condition
\begin{equation}
\label{trivial}
z' = \xi , \qquad \xi \equiv  
\frac{\beta -1}{2 \sqrt{\beta}} , \qquad \beta \ne \{0, 1 \}  \, .
\end{equation}
Hence, the general solution to Eq. (\ref{I2I4ansatzode}) is given by 
\begin{equation}
\label{gensolz}
I_2 (\phi ) = \left( \frac{2C}{Q} \right)^{1/(\beta -1)} [ {\rm sech} \, 
\xi (\phi -\phi_0) ]^{2/(\beta -1)} \, .
\end{equation}
Since $\phi$ is a monotonically varying function of time, the solution 
asymptotes from $I_2 (-\infty )$ to $I_2 (+\infty )$. This 
represents a configuration that is initially expanding, reaches a point 
of maximal expansion, and then proceeds to recollapse.   

To proceed, it is natural 
to consider the nature of the solution when 
$\beta =1$ in {\em ansatz} (\ref{I2I4ansatz}). 
It is clear from Eq. (\ref{I2only}) that $I_2$ depends exponentially on the 
scalar field in this case. 
To understand the physical significance of this solution, 
we first consider a specific class of cosmological 
solutions in the following Subsection. 

\subsection{Scaling Solutions}

One class of cosmological solutions that are of particular 
interest are `scaling solutions', whereby the energy densities of the 
scalar field and perfect fluid vary at the same rate as the universe 
evolves. (See, e.g., Ref. \cite{Ed} for an exhaustive list of 
references). Indeed, these solutions describe a universe 
that expands (or contracts) as if it were sourced only by 
the perfect fluid and the scalar field is said to `track' the fluid. 
These solutions are important 
because they typically represent early- or late-time 
attractors and repellors (critical points in the phase space) 
for more general solutions and they thereby 
allow one to determine the asymptotic 
behaviour and stability of more complicated 
models. 

Cosmological scaling solutions
are characterized by the condition $\Omega_{\phi} = {\rm constant}$ with 
$0 <n \le 6$. This is equivalent to the conditions 
$I_3={\rm constant}$ and $I_2I_4 = {\rm constant}$, i.e., to 
{\em ansatz} (\ref{I2I4ansatz}) with $\beta =1$. It follows 
from Eq. (\ref{HJQconstraint}) that $I_2'I_4'$ is also $\phi$-invariant. 
This implies that $I_2(\phi)$ and $I_4(\phi)$ are both 
purely exponential functions 
of the scalar field. Consequently, the integration constant in solution 
(\ref{I2quad}) must vanish, $\Pi_2 =0$.
The wavefunction analogue of the 
cosmological scaling solution is therefore readily deduced
from Eqs. (\ref{invariantequation}) and (\ref{HJQconstraint}) 
and we find that 
\begin{equation}
I_2 (\phi) = 2 \left( \frac{Q}{\Omega_{\phi}-1 } \right)^{1/2} 
e^{\sqrt{n/\Omega_{\phi}}\phi} , 
\qquad  I_4 (\phi) = \frac{\Omega_{\phi}}{4} 
\left( \frac{Q}{\Omega_{\phi} -1} 
\right)^{1/2} e^{-\sqrt{n/\Omega_{\phi}}\phi}  \, .
\end{equation}
Laboratory time is related to the value of the scalar field via 
$\phi = \sqrt{\Omega_{\phi}/n} \ln (\tau -\tau_0 )$, where $\tau_0$ is an
arbitrary constant. Hence, 
 \begin{equation}
\label{condensatescaling}
I_2 (\tau) = 2 \left( \frac{Q}{\Omega_{\phi}-1} \right)^{1/2}
(\tau -\tau_0), \qquad 
I_4 (\tau ) = \frac{\Omega_{\phi}}{4} \left( \frac{Q}{\Omega_{\phi} -1} 
\right)^{1/2} \frac{1}{\tau -\tau_0}, 
\qquad \lambda (\tau) = \frac{\Omega_{\phi}}{4} 
\frac{1}{(\tau -\tau_0 )^2} \, .
\end{equation}

Finally, we note that 
the dual solution generated from the form-invariance transformation 
of Section \ref{HJNLS} is given by
 \begin{equation}
\label{dualcondensatescaling}
\tilde{I}_2 (\tilde{\tau}) = 2 \left( \frac{Q}{\Omega_{\phi}-1} \right)^{1/2}
(\tilde{\tau} -\tilde{\tau}_0), \qquad 
\tilde{I}_4 (\tilde{\tau} ) = \frac{\Omega_{\phi}}{4} \left( 
\frac{Q}{\Omega_{\phi} -1} 
\right)^{1/2} \frac{1}{\tilde{\tau} -\tilde{\tau}_0}, 
\qquad \tilde{\lambda} (\tilde{\tau}) = \frac{\Omega_{\phi}}{4} 
\frac{1}{(\tilde{\tau} -\tilde{\tau}_0 )^2}  \, ,
\end{equation}
where the time-parameters are related by 
$\tilde{\tau} - \tilde{\tau}_0 =
\Omega_{\phi}/[8 (\tau-\tau_0)]$. In this sense, the scaling solution is 
the self-dual solution of the system 
(\ref{I4phiode})-(\ref{HJQconstraint}). 

\section{Conclusion and Discussion}

\label{conclusion}

\def\theequation{\arabic{equation}}

In this paper, we have established a direct link 
between the dynamics of two apparently disparate systems, namely between
the dynamics arising 
from the radially-symmetric, cubic Schr\"odinger equation 
and the cosmological Friedmann
equations sourced by a self-interacting scalar field and barotropic perfect
fluid. The correspondence is established between the moments of the 
wavefunction and the cosmological parameters. Whilst 
similar analogies between 
the physical variables of the two systems have been 
considered previously in the literature for spatially curved and anisotropic 
Bianchi I models
\cite{lidseycondensate,ambroise,ambroisePhD}, 
we have made the correspondence
precise at the level of the equations of motion (field equations). This 
correspondence is summarized in Tables I and II. In effect, the dictionary 
between the two systems is equivalent to a change of dependent 
variables and a reparametrization of cosmic and laboratory times. 

By analogy with the Hamilton-Jacobi formalism of 
scalar field cosmology, we have expressed the moment evolution 
equations in an alternative, first-order form. This 
uncovered a form-invariance in the ODEs, 
generally relating expanding configurations to
contracting ones. It also allows parametric solutions to the EP 
equation to be found and we introduced 
algorithms for finding solutions with some worked examples. 
We emphasize that for consistency, 
the scalar field must be a monotonically varying function 
of laboratory (cosmic) time in this formalism. 
Consequently, the trapping potential 
frequency $\lambda (\tau) $ can not pass through zero. 

Whether such a correspondence 
between a nonlinear quantum system and the classical 
Einstein equations is more than an
intriguing mathematical analogy (or indeed no more than 
a coincidence) remains to 
be seen. It would be interesting to explore 
this issue further. One question that immediately arises is whether 
the correspondence can be applied to the more general $d$-dimensional 
NLS equation (\ref{genNLS}) when $d \ne 2$. 
In the absence of a trapping potential 
($V=0$), the nonlinear interacting theory is defined by the action
\begin{equation}
S= \int d\tau \int d^dx \left[ \frac{i}{2}  \left( 
u^* \frac{\partial u}{\partial \tau} - u \frac{\partial u^*}{\partial \tau}
\right) -\frac{1}{2} \nabla u^* \cdot \nabla u - \frac{\nu}{m}
(u^*u)^m \right]
\end{equation}
and the free field theory $(\nu =0)$ is a one-dimensional, 
non-relativistic conformal 
field theory exhibiting a dynamical $SO(2,1) \cong SL(2,R)$
symmetry \cite{conformal,conformal1}. 
In general, the $U(1)$-invariant polynomial
interaction term breaks the scale-invariance of the free theory unless
the condition 
\begin{equation}
\label{conformal}
md =d+2
\end{equation}
is satisfied, i.e., the coupling constant $\nu$ is dimensionless. 

On the other hand, 
for arbitrary $d$, the moments of the wavefunction are defined by 
\begin{eqnarray}
I_2 & = & \int |u|^2 r^2 d^dx \, ,
\\
I_3 & = & i \int  \left( u \frac{\partial u^*}{\partial r} -u^* 
\frac{\partial u}{\partial r} \right) r d^dx \, ,
\\
I_4 & = & 
\int \left(  \frac{1}{2} \left| \nabla u \right|^2 + 
\frac{\nu}{m}  (uu^*)^m \right)  d^d x \, .
\end{eqnarray}
By repeating the analysis of section \ref{bose-einstein}, it 
can be shown that the time-evolution of these moments 
is determined by the set of equations
\begin{eqnarray}
\frac{dI_2}{d\tau} & = & I_3 \, ,
\\
\frac{dI_3}{d\tau} & = & -2 \lambda (\tau ) I_2 
+4 \int d^d x \left( \frac{1}{2} | \nabla u |^2 + \nu \frac{d(m-1)}{2m}
|u|^{2m} \right) \, ,
\\
\frac{dI_4}{d\tau } & = & -\frac{1}{2} \lambda (\tau ) I_3 \, .
\end{eqnarray}
It follows that the system reduces to that of the 
two-dimensional case (\ref{I2ode})-(\ref{I4ode})
{\em if} the conformal condition (\ref{conformal}) is satisfied. 
%This requires $m=3$ in the case $d=1$, and $m=2$ for $d=2$. 
%As $d \rightarrow \infty$, we have $m \rightarrow 1$. 
In this case, the moment evolution equations can be identified with the 
cosmological Friedmann equations as for the two-dimensional NLS equation. 
At this level, therefore, 
the correspondence seems to be associated with non-relativistic 
conformal field theory and it would be interesting to explore this possibility
further.  

In conclusion, it is well known 
that universes sourced by scalar fields and 
perfect barotropic fluids have played a central role in modern 
cosmology over recent decades. We have found that such models also 
have direct applications in studies of nonlinear quantum mechanics.

\section*{Acknowledgements}

We thank R. Tavakol for numerous helpful discussions.

\section*{APPENDIX: A WORKED EXAMPLE}

\def\theequation{\arabic{equation}}

As a worked example of the solution-generating 
techniques developed in Section \ref{algorithms}, 
let us consider the {\em ansatz}
\begin{equation}
\label{ansatz1}
I_4(\phi) = A \sinh \, \omega \phi \, ,
\end{equation}
where $A$ is an arbitrary constant and $n \equiv -\omega^2$. 
It follows after integrating Eq. (\ref{I2quad}) 
that the (square) width is given by  
\begin{equation} 
I_2(\phi) = \Pi_2 \cosh \, \omega\phi -\frac{Q}{2A} \sinh \omega\phi
\end{equation}
and laboratory time is determined from Eq. (\ref{tauphi}) in terms 
of the quadrature   
\begin{equation}
\label{ansatz1tauphi}
\tau (\phi) = \frac{n}{4A\omega} \int d\phi \sqrt{2A \Pi_2 {\rm tanh} \, 
\omega\phi -Q} \, .
\end{equation}
Although the integral (\ref{ansatz1tauphi}) can be evaluated analytically, 
the result is not invertible in general. To proceed, therefore,
we specify $|Q| = 2 A \Pi_2$, 
where the modulus sign applies if $Q<0$. It then follows that 
\begin{equation}
I_2(\phi) = \Pi_2 e^{\omega \phi}
\end{equation}
and the corresponding expression for the potential frequency derives from 
Eq. (\ref{lambdaphi}): 
\begin{equation}
\lambda (\phi) = -\frac{4A^2}{|Q|} \frac{1}{1+{\rm tanh} \, \omega \phi}
\, .
\end{equation}

For illustrative purposes, 
let us consider the case $Q<0$. We may then employ the 
standard integral given by  
\begin{equation}
\label{negQintegral}
\int d \phi \sqrt{1+ {\rm tanh} \, \omega\phi} =  
\frac{\sqrt{2}}{\omega}
{\rm tanh}^{-1}\, \left( \sqrt{\frac{1+ {\rm tanh} 
\, \omega \phi}{2}} \right)
\end{equation}
to deduce from Eq. (\ref{ansatz1tauphi}) that 
\begin{eqnarray} 
\label{sinh}
\omega \phi & = & {\rm tanh}^{-1} \left[ 
2 {\rm tanh}^2 \left( \sqrt{\frac{8A^2}{|Q|}} (\tau 
-\tau_0) \right) -1 \right] \\
& = & \ln \sinh \left( \sqrt{\frac{8A^2}{|Q|}} (\tau -\tau_0) \right) \, ,
\end{eqnarray}
where $\tau_0$ is the arbitrary integration constant. 
Substituting Eq. (\ref{sinh}) into the relevant expressions for the 
moments then leads (after appropriate trigonometric 
identities have been employed) to the full solution
\begin{eqnarray}
\lambda (\tau) & = &
- \frac{2A^2}{|Q|} {\rm cotanh}^2 \, \left( \sqrt{\frac{8A^2}{|Q|}}
(\tau - \tau_0) \right) \, ,
\\
I_2 (\tau) & = & X^2(\tau ) = \Pi_2 \sinh \, 
\left( \sqrt{\frac{8A^2}{|Q|}}
(\tau -\tau_0)  \right) \, ,
\\
I_3(\tau) & = & \sqrt{2|Q|} \cosh \left( \sqrt{\frac{8A^2}{|Q|}}
\tau -\tau_0 ) \right) \, ,
\\
I_4(\tau ) & = & \frac{A}{2} {\rm cosech} \left( \sqrt{\frac{8A^2}{|Q|}}
\tau \right) \left[ \cosh^2 \left( \sqrt{\frac{8A^2}{|Q|}}
(\tau -\tau_0) \right) -2 \right] \, .
\end{eqnarray}


\begin{thebibliography}{99}

\bibitem{G}
E. P. Gross, Nuovo Cimento {\bf 20}, 454 (1961).

\bibitem{P}
L. P. Pitaevskii, Zh. Eksp. Teor. Fiz. {\bf 40}, 646 (1961)
[Sov. Phys. JETP {\bf 13}, 451 (1961)]. 

\bibitem{volovik1}
G. E. Volovik, Phys. Rep. {\bf 351}, 195 (2001). 

\bibitem{volovik2}
G. E. Volovik, {\em The Universe in a Helium Droplet} (Oxford
University Press, Oxford, 2003). 

\bibitem{frant}
D. J. Frantzeskakis, 	
J. Phys. A {\bf 43}, 213001 (2010).

\bibitem{hasegawa}
A. Hasegawa, {\em Optical Solitons in Fibres}
(Springer-Verlag, Berlin, 1990).

\bibitem{moment}
S. N. Vlasov, V. A. Petrishchev, and V. I. Talanov, Radiophys. Quantum
Electron. {\bf 14}, 1062 (1971); 
M. A. Porras, J. Alda and E. Bernabeu, 
Appl. Opt. {\bf 32}, 5885 (1993).

\bibitem{moment1}
J. J. Garcia-Ripoll, V. M. Perez-Garcia, and P. Torres, 
Phys. Rev. Lett. {\bf 83}, 1715 (1999). 

\bibitem{moment2}
V. M. Perez-Garcia, M. A. Porras, and L. Vazquez, 
Phys. Lett. A {\bf 202}, 176 (1995). 

\bibitem{moment3}
J. J. Garcia-Ripoll and V. M. Perez-Garcia, patt-sol/9904006. 

\bibitem{moment4} 
G. D. Montesinos, V. M. Perez-Garcia, and P. J. Torres, 
Physica D {\bf  191}, 193 (2004).

\bibitem{lidsey}
J. E. Lidsey, A. R. Liddle, E. W. Kolb, E. J. Copeland, 
T. Barreiro, and M. Abney, 
{\em Rev. Mod. Phys.} {\bf 69}, 373 (1997). 

\bibitem{lr}
D. H. Lyth and A. Riotto, Phys. Rep. {\bf 314}, 1 (1999). 

\bibitem{cst}
E. J. Copeland, M. Sami, and S. Tsujikawa, Int. J. Mod. Phys. D {\bf 15}, 1753
(2006). 

\bibitem{clifton}
T. Clifton, P. G. Ferreira, A. Padilla, and C. Skordis,
Phys. Rep. {\bf 513}, 1 (2012).

\bibitem{pbb1}
M. Gasperini and  G. Veneziano, Astropart. Phys. {\bf 1}, 317 (1993). 

\bibitem{pbb2}
J. E. Lidsey, D. Wands, and E. J. Copeland, Phys. Rep. {\bf 337}, 343 (2000).

\bibitem{ermakov}
V. P. Ermakov, Univ. Izv. Kiev {\bf 20}, 1 (1880). 

\bibitem{pinney}
E. Pinney, Proc. Am. Math. Soc. {\bf 1}, 681 (1950).

\bibitem{milne}
W. E. Milne, Phys. Rev. {\bf 35}, 863 (1930). 

\bibitem{EPsurvey}
P. G. L. Leach and K. Andriopoulos, Appl. Anal. Discrete Math. {\bf 2}, 146
(2008). 

\bibitem{EPsurvey1}
V. P. Akulov, S. Catto, O. Cebecioglu, and A. Pashnev, 
Phys. Lett. B {\bf 575}, 137 (2003). 

\bibitem{hawklid}
R. M. Hawkins and J. E. Lidsey, Phys. Rev. D {\bf 66}, 023523 (2002). 

\bibitem{williams}
F. L. Williams and P. G. Kevrekidis, 
Class. Quantum Grav. {\bf 20}, L177 (2003).

\bibitem{lidseycondensate}
J. E. Lidsey, Class. Quantum Grav. {\bf 21}, 1 (2004). 

\bibitem{williams1}
F. L. Williams, P. G. Kevrekidis, T. Christodoulakis, C. Helias,  
G. O. Papadopoulos, and Th. Grammenos, 
gr-qc/0408056.

\bibitem{herring}
G. Herring, P. G. Kevrekidis, F. Williams, 
T. Christodoulakis, and D. J. Frantzeskakis,
Phys. Lett. A {\bf 367}, 140 (2007).

\bibitem{ambroise}
J. D'Ambroise and F. L. Williams, 
J. Math. Phys. {\bf 51}, 062501 (2010).

\bibitem{ambroisePhD}
J. D'Ambroise, {\em PhD. Thesis}, arXiv:1005.1410. 
%Generalized EMP and nonlinear Schrodinger-type 
%reformulations of some scalar field cosmological models
     
\bibitem{HJ}
L. P. Grishchuk and Yu. V. Sidorav, in {\em Fourth 
Seminar on Quantum Gravity}, eds. M. A. Markov, V. A. Berezin, and V. P. Frolov
(World Scientific, Singapore, 1988). 

\bibitem{HJ1}
D. S. Salopek, J. R. Bond, and J. M. Bardeen, 
Phys. Rev. D {\bf 40}, 1753 (1989).

\bibitem{HJ2}
 A. G. Muslimov, Class. Quantum Grav. {\bf 7}, 231 (1990). 

\bibitem{HJ3} 
D. S. Salopek and J. R. Bond, Phys. Rev. D {\bf 42}, 3936 (1990). 

\bibitem{HJ4}  
B. J. Carr and J. E. Lidsey, Phys. Rev. D {\bf 48}, 543 (1990). 

\bibitem{lidsey*}
J. E. Lidsey, Class. Quantum Grav. {\bf 8}, 923 (1991). 

\bibitem{lidsey1}
J. E. Lidsey, Phys. Lett. B {\bf 273}, 42 (1991).

\bibitem{chimento}
L. P. Chimento, 
J. Math. Phys. {\bf 38}, 2565 (1997).

\bibitem{sahni}
Y. Shtanov and V. Sahni, Phys. Lett. B {\bf 557}, 1 (2003). 

\bibitem{param}
P. Singh, K. Vandersloot, and G. V. Vereshchagin, 
Phys. Rev. D {\bf 74}, 043510 (2006). 

\bibitem{lidsey2}
J. E. Lidsey, Phys. Rev. D {\bf 70}, 041302 (2004). 

\bibitem{Ed}
E. J. Copeland, S. Mizuno, and M. Shaeri, 
Phys. Rev. D {\bf 79}, 103515 (2009). 

\bibitem{conformal}
U. Niederer, Helv. Phys. Acta. {\bf 45}, 802 (1972).

\bibitem{conformal1}
R. Jackiw and S.-Y. Pi, Phys. Rev. D {\bf 42}, 3500 (1990). 


\end{thebibliography}
\end{document}